\documentclass{article}
\usepackage{ijcai19}

\usepackage{times}
\usepackage{soul}
\usepackage{url}
\usepackage[hidelinks]{hyperref}
\usepackage[utf8]{inputenc}
\usepackage[small]{caption}
\usepackage{graphicx}
\usepackage{amsmath}
\usepackage{booktabs}
\usepackage{subcaption}
\usepackage{multirow}
\usepackage{comment}

\usepackage{color}
\makeatletter
\let\blx@rerun@biber\relax
\makeatother

\title{Musical Composition Style Transfer via Disentangled Timbre Representations}

\author{
Yun-Ning Hung$^1$\and 
I-Tung Chiang$^1$\and
Yi-An Chen$^2$\and
Yi-Hsuan Yang$^1$\\
\affiliations
$^1$Research Center for IT Innovation, Academia Sinica, Taiwan\\
$^2$KKBOX Inc., Taiwan\\
\emails
\{biboamy, itungchiang\}@citi.sinica.edu,
annchen@kkbox.com,
yang@citi.sinica.edu
}

\begin{document}

\maketitle

\begin{abstract}
Music creation involves not only composing the different parts (e.g., melody, chords) of a musical work but also arranging/selecting the instruments to play the different parts.
While the former has received increasing attention, the latter has not been much investigated.
This paper presents, to the best of our knowledge, the first deep learning models for rearranging music of arbitrary genres. Specifically, we build encoders and decoders that take a piece of polyphonic musical audio as input, and predict as output its musical score. We investigate disentanglement techniques such as adversarial training to separate latent factors that are related to the musical content (pitch) of different parts of the  piece, and that are related to the instrumentation (timbre) of the parts 
per short-time segment. 
By disentangling pitch and timbre, our models have an idea of how each piece was composed and arranged. Moreover, the models can realize ``composition style transfer'' by rearranging a musical piece without much affecting its pitch content. 
We validate the effectiveness of the models by experiments on instrument activity detection and composition style transfer.
To facilitate follow-up research, we open source our code at  \url{https://github.com/biboamy/instrument-disentangle}.
\end{abstract}

\section{Introduction}

Music generation has long been considered as an important task for AI, 
possibly due to the complicated mental and creative process that is involved, and its wide applications in our daily lives. Even if machines cannot reach the same professional level as well-trained composers or songwriters, machines can cooperate with human to make music generation easier or more accessible to everyone.

Along with the rapid development of deep generative models such as 
generative adversarial networks (GANs) \cite{goodfellow2014generative}, 
the research on music generation is experiencing a new momentum. 
While a musical work is typically composed of melody, chord, and rhythm (beats), a major trend of recent research focuses on creating original content for all or some of these three parts \cite{roberts17nips,dong2018musegan}.
That is, we expect machines can learn the complex musical structure and compose music like human composers. This has been technically referred to as \emph{music composition} \cite{briot17survey}. 

Another trend is to generate variations of existing musical material. By preserving the pitch content of a musical piece, we expect machines to modify style-related attributes of music, such as genre and instrumentation. This has been technically referred to as \emph{music style transfer} \cite{dai2018music}. A famous example of style transfer, though not fully-automatic, is the ``Ode to Joy'' project presented by \cite{pachet16tist}. 
The project aims to rearrange (or reorchestrate) Beethoven’s Ode to Joy according to the stylistic conventions of seven other different musical genres such as Jazz and Bossa Nova.\footnote{[Online]~\url{https://www.youtube.com/watch?v=buXqNqBFd6E}}


In this paper, we are in particular interested in such a \emph{music rearrangement} task. The work of \cite{pachet16tist} is inspiring, but as the target styles are by nature very different, they had to tailor different machine learning models, one for each style. 
Moreover, they chose to use non-neural network based models such as Markov models, possibly due to the need to inject musical knowledge manually to ensure the final production quality.
As a result, the solution presented by \cite{pachet16tist} is not general and it is not easy to make extensions.
In contrast, we would like to investigate a universal neural network model with which we can easily change the instrumentation of a given music piece fully automatically, without much affecting its pitch content. While we do not think such a style-agnostic approach can work well for all the possible musical styles, it can be treated as a starting point from which style-specific elements can be added later on.

Music rearrangement is closely related to the art of \emph{music arrangement} \cite{arrangement}. For creating music, it involves not only composing the different parts (e.g., melody, chords) of a musical work  but also arranging/selecting the  instruments to play the different parts. 
For example, chord and melody can be played by different instruments. 
Setting the instrumentation is like \emph{coloring} the music.
Therefore, music arrangement is an ``advanced'' topic
, and is needed towards generating fully-fledged music material.
Compared to music arrangement, music rearrangement is a style transfer task and is like \emph{re-coloring} an existing music piece.
By contributing to music rearrangement, we also contribute indirectly to the more challenging 
automatic music arrangement task, which has been seldom addressed in the literature as well.

Music rearrangement is not an easy task even for human beings and requires years of professional training. 
In order to convincingly perform music style transfer between two specific styles, a musician must first be familiar with the annotation, instrument usage, tonality, harmonization and theories of the two styles. The task gets harder when we consider more styles.
For machines to rearrange a musical piece, we have to consider not only the pitch content of the given music, but also the pitch range of each individual instrument as well as the relation between different instruments (e.g., some instrument combination creates more harmonic sounds). 
And, it is hard to find \emph{paired data} demonstrating different instrumentations of the same music pieces \cite{Crestel2017ADL}. 
A neural network may not learn well when the data size is small.


To address these challenges, we propose to train \emph{music transcription} neural networks that take as input  an audio file and generate as output the corresponding musical score. 
In addition to predicting which notes are played in the given audio file, the model also has to predict the instrument that plays each note. As a result, the model learns both pitch- and timbre-related representations from the musical audio.  While the model is not trained to perform music rearrangement, we investigate ``disentanglement techniques'' such as adversarial training \cite{liu2018exploring,lee2018diverse} to separate latent factors of the model that are pitch- and timbre-related. 
By disentangling pitch and timbre, the model has an idea of how a musical piece was composed and arranged.
If the disentanglement is well done, we expect that we can rearrange a musical piece by holding its pitch-related latent factors fixed while manipulating the timbre-related ones.

We propose and investigate two models in this paper. They share the following two advantages. First, they require only pairs of audio and MIDI files, which are relatively easier to collect than different arrangements of the same music pieces. 
Second, they can take any musical piece as input and rearrange it, as long as we have its audio recording.



In sum, the main contributions of this work include:
\begin{itemize}
\item To the best of our knowledge, this work presents the first deep learning models that realize music rearrangement, a composition style transfer task \cite{dai2018music} (see Section \ref{sec:relatedwork1}), for polyphonic music of arbitrary genres. 
\item While visual attribute disentanglement has been much studied recently, attribute disentanglement for musical audio has only been studied in our prior work \cite{hung2018learning}, to our knowledge. Extending from that work, we report comprehensive evaluation of the models, emphasizing  their application to  rearrangement.
\end{itemize}
Musical compositional style transfer is still at its early stage. We hope this work can call for more attention toward approaching this task with computational means.



\section{Related Work}
\label{sec:relatedwork}

\subsection{Music Style Transfer}
\label{sec:relatedwork1}
According to the categorization of \cite{dai2018music}, there are three types of music style transfer: \emph{timbre style transfer}, \emph{performance style transfer}, and \emph{composition style transfer}. 
To our knowledge, timbre style transfer receives more attention than the other two lately. The task is concerned with altering the timbre of a musical piece in the audio domain. For example, \cite{engel2017neural} applied WaveNet-based autoencoders to audio waveforms to model the musical timbre of short single notes, and \cite{lu2018play} employed a GAN-based network on spectrograms to achieve long-structure timbre style transfer.
However, for the latter two types of music style transfer, little effort has been made. 

Music rearrangement can be considered as a composition style transfer task, which is defined as a task that ``preserve[s] the identifiable melody contour (and the underlying structural functions of harmony) while altering some other score features in a meaningful way'' \cite{dai2018music}. 
The recent work on composition style transfer we are aware of are that by
\cite{Pati} and \cite{kaliakatsos2017conceptual}, which used neural networks and rules respectively to deal with melody, rhythm, and chord modification. 
Our work is different in that we aim to modify the instrumentation. 

\subsection{Disentanglement for Style Transfer}
\label{sec:relatedwork2}
Neural style transfer was first introduced in image style transfer \cite{gatys2016image}. In this work, they proposed to learn an image representation to separate and recombine  content and style. Since then, several methods have been proposed to modify the style of a image. \cite{liu2018exploring} trained an auto-encoder model supervised by content ground truth labels to learn a content-invariant representation, which can be modified to reconstruct a new image. 
\cite{lee2018diverse} showed that it is possible to learn a disentangled representation from unpaired data by using cross-cycle consistency loss and adversarial training, without needing the ground truth labels.


Disentangled representation learning for music is still new and has only been studied by \cite{brunner2018midi} for symbolic data. 
By learning a style representation through style classifier, their model is able to modify the pitch, instrumentation and velocity given different style codes. 
Our model is different in that we deal with audio input rather than MIDIs. 


\begin{figure}[t] 
\centering
\begin{subfigure}[b]{0.5\textwidth} 
\centering
\includegraphics[width=\textwidth]{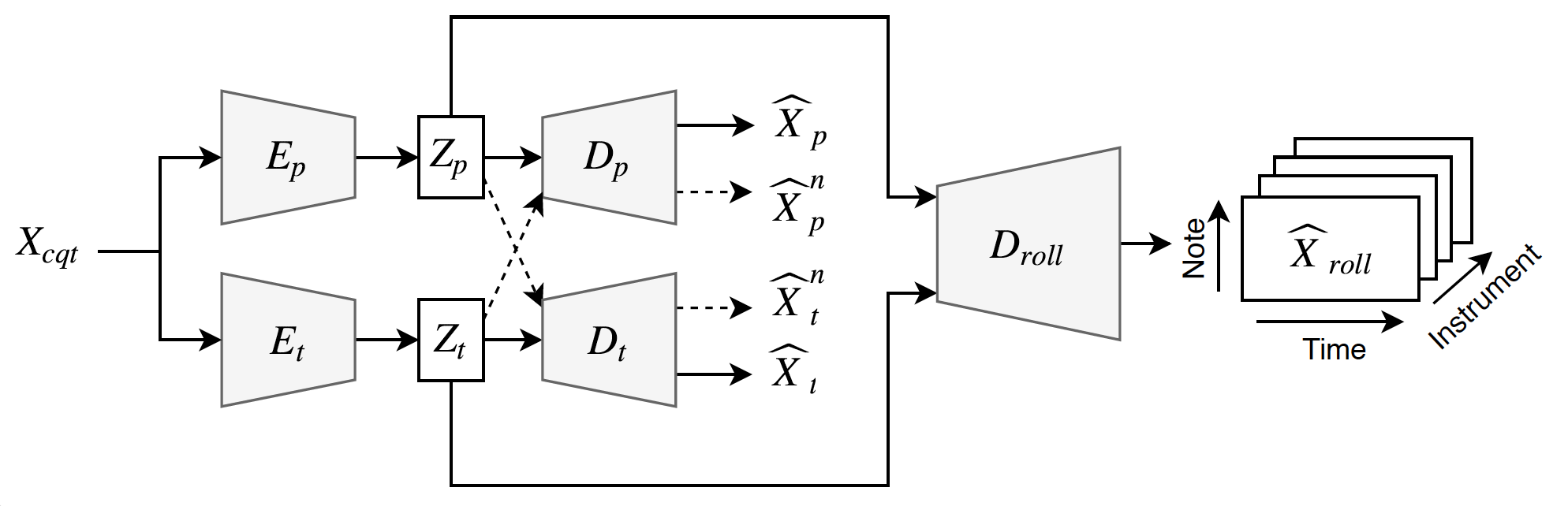}
\caption{DuoED: Use separate encoders/decoders for timbre and pitch}
\label{fig: duoae}
\vspace{3mm}
\centering
\includegraphics[width=0.8\textwidth]{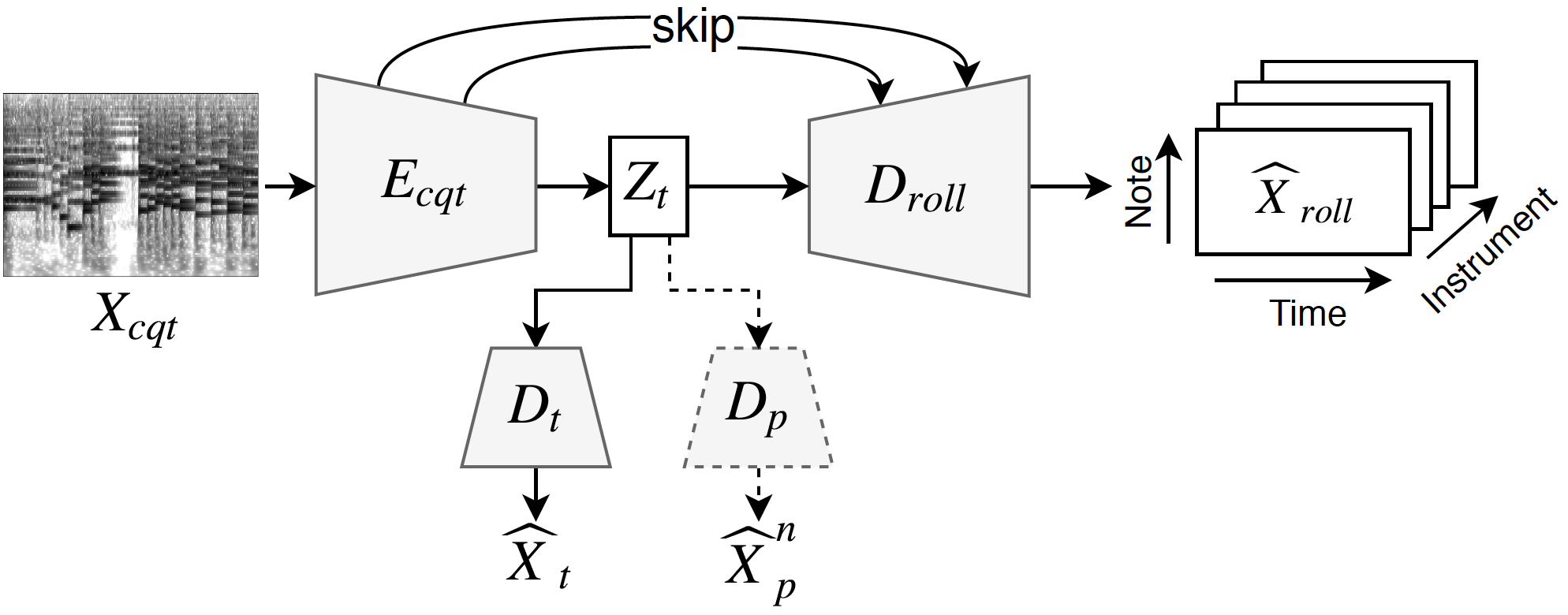}
\caption{UnetED: Use skip connections to process pitch}
\label{fig: unetae}
\end{subfigure}
\caption{The two proposed encoder/decoder architectures for learning disentangled timbre representations (i.e., $\mathbf{Z}_t$) for musical audio. The dashed lines indicate the adversarial training components. (Notations: CQT---a time-frequency representation of audio; roll--multi-track pianoroll; E---encoder; D---decoder; Z---latent code; t---timbre; p---pitch; skip---skip connections).}
\label{fig: model}
\end{figure}

\begin{figure}
\centering
\begin{subfigure}[b]{0.17\textwidth}
\includegraphics[width=\textwidth]{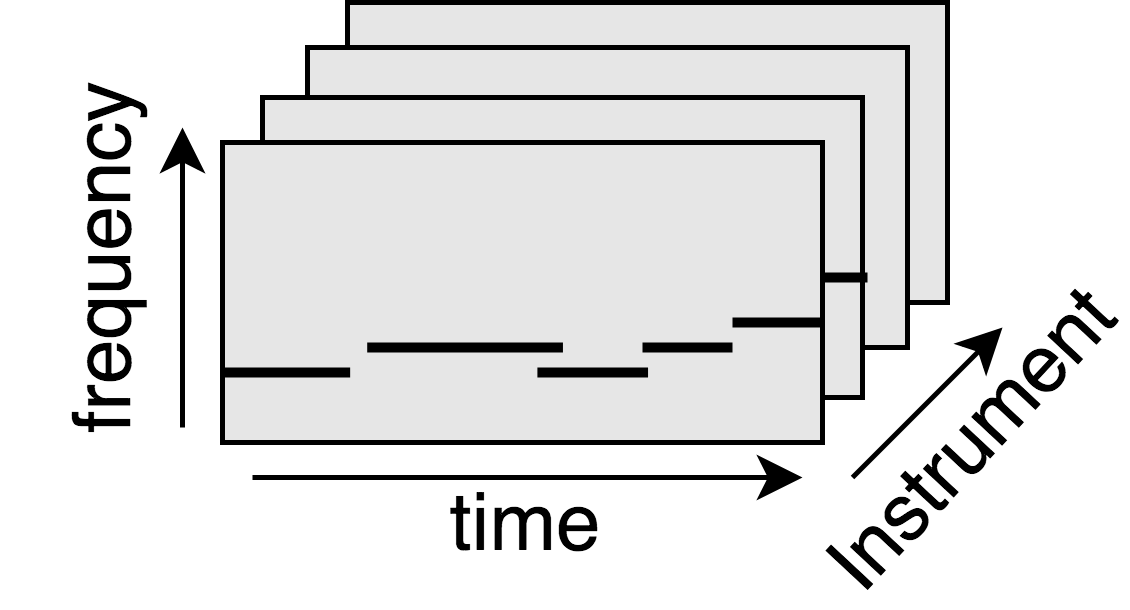}
\caption{Pianoroll}
\label{fig: pianoroll}
\end{subfigure}
\begin{subfigure}[b]{0.14\textwidth}
\includegraphics[width=\textwidth]{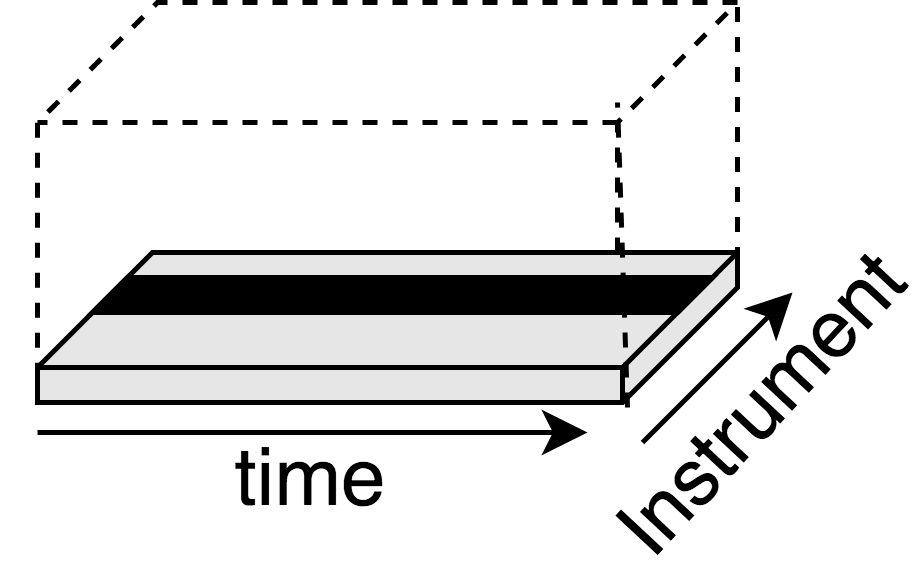}
\caption{Instrument roll}
\label{fig: inst_roll}
\end{subfigure}
\begin{subfigure}[b]{0.14\textwidth}
\includegraphics[width=\textwidth]{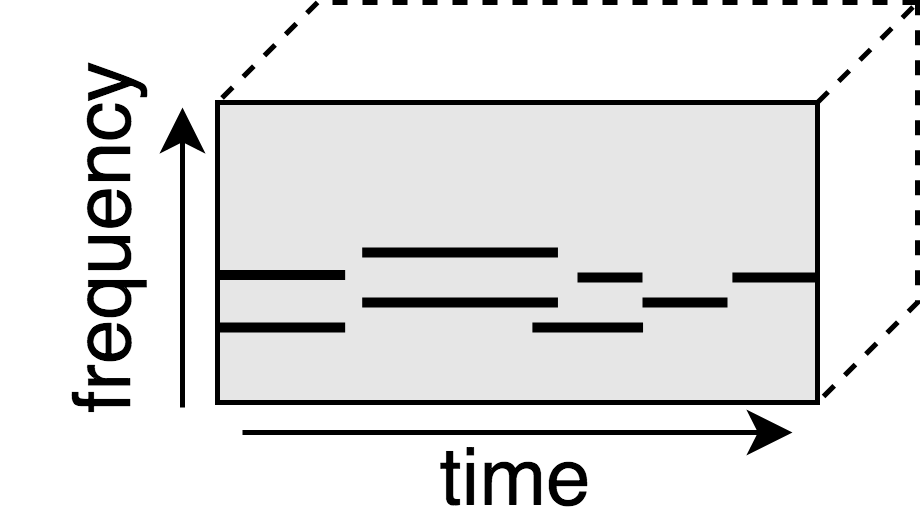}
\caption{Pitch roll}
\label{fig: pitch_roll}
\end{subfigure}
\caption{Different symbolic representations of music.}
\label{fig: roll}
\end{figure}

\section{Proposed Models}
\label{sec:model}

Figure \ref{fig: model} shows the architecture of the proposed models. 
We opt for using the encoder/decoder structures as the backbone of our models, as such networks learn latent representations (a.k.a., latent codes) of the input data.
With some labeled data, it is possible to train an encoder/decoder network in a way that different parts of the latent code correspond to different data attributes, such as pitch and timbre. 
Both models
use encoders/decoders and adversarial training to learn music
representations, but the second model additionally uses skip
connections to deal with the pitch information.
We note that both models perform polyphonic music transcription.
We present the details of these models below.

\subsection{Input/Output Data Representation}

The input to our models is an audio waveform with arbitrary length. To facilitate pitch and timbre analysis, we firstly convert the waveform into a time-frequency representation that shows the energy distribution across different frequency bins for each short-time frame. Instead of using the short-time Fourier transform, we use the constant-Q transform (CQT), for it adopts a logarithmic frequency scale that better aligns with our perception of pitch~\cite{bittner17ismir}. CQT also provides better frequency resolution in the low-frequency part, which helps detect the fundamental frequencies.

As will be described in Section \ref{sec:implementation_details}, 
our encoders and decoders are designed to be fully-convolutional~\cite{oquab15localization,liu2018weakly}, so that our models can deal with input of any length in testing time. However, for the convenience of training the models with mini-batches, in the training stage we divide the waveforms in our training set into 10-second chunks (without overlaps) and use these chunks as the model input, leading to a matrix $\mathbf{X}_{cqt} \in \mathcal{R}^{F \times T}$ of fixed size for each input.
In our implementation, we compute CQT with the \texttt{librosa} library~\cite{librosa}, with 16,000 Hz sampling rate and 512-sample window size, again with no overlaps. We use a frequency scale of 88 bins, with 12 bins per octave to represent each note. Hence, $F=88$ (bins) and $T=312$ (frames).

We use the pianorolls \cite{pypianoroll} (see Figure \ref{fig: pianoroll})
as the target output of our models. A pianoroll is a binary-valued tensor that records the presence of notes (88 notes here) across time for each track (i.e., instrument). When we consider $M$ instruments, the target model output would be $\mathbf{X}_{roll} \in \{0,1\}^{F \times T \times M}$. $\mathbf{X}_{roll}$ and $\mathbf{X}_{cqt}$ are temporally aligned, since we use MIDIs that are time-aligned with the audio clips to derive the pianorolls, as discussed in Section \ref{sec:implementation_details}. As shown in Figures \ref{fig: model} and \ref{fig: roll}, besides asking our models to generate $\mathbf{X}_{roll}$, we use the \emph{instrument roll} $\mathbf{X}_{t} \in \{0,1\}^{M \times T}$ and \emph{pitch roll} $\mathbf{X}_{p} \in \{0,1\}^{F \times T}$ as supervisory signals to help learn the timbre representation.  Both $\mathbf{X}_{t}$ and $\mathbf{X}_{p}$ can be computed from $\mathbf{X}_{roll}$ by summing along a certain dimension.

\subsection{The DuoED Model}
The architecture of DuoED is illustrated in Figure \ref{fig: model}a. The design is inspired by~\cite{liu2018exploring}, but we 
adapt the model to encode music. 
Specifically, we train two encoders $E_t$ and $E_p$ to respectively convert $\mathbf{X}_{cqt}$ into the \emph{timbre code} $\mathbf{Z}_t = E_t(\mathbf{X}_{cqt}) \in \mathcal{R}^{\kappa \times \tau}$ and \emph{pitch code} $\mathbf{Z}_p = E_p(\mathbf{X}_{cqt}) \in \mathcal{R}^{\kappa \times \tau}$.
We note that, unlike in the case of image representation learning, here the latent codes are \emph{matrices}, and we require that the second dimensions (i.e., $\tau$) represent time. This way, each column of $\mathbf{Z}_t$ and $\mathbf{Z}_p$ is a $\kappa$-dimensional representation of a temporal segment of the input. For abstraction, we require $\kappa\tau < FT$. 

DuoED also contains three decoders $D_{roll}$, $D_t$ and $D_p$.  The encoders and decoders are trained such that we can use $D_{roll}([\mathbf{Z}_t^T, \mathbf{Z}_p^T]^T)$ to predict $\mathbf{X}_{roll}$,  $D_t(\mathbf{Z}_t)$ to predict $\mathbf{X}_t$, and  $D_p(\mathbf{Z}_p)$ to predict $\mathbf{X}_p$. The prediction error is measured by the \emph{binary cross entropy} between the ground truth and the predicted one. For example, for the timbre classifier $D_t$, it is:
\begin{equation} 
L_t = - \textstyle{\sum} ~[\mathbf{X}_t\cdot \ln\sigma(\widehat{\mathbf{X}}_t) + (1-\mathbf{X}_t) \cdot \ln(1-\sigma(\widehat{\mathbf{X}}_t)) ] \,,
\label{eq: BCN}
\end{equation}
where $\widehat{\mathbf{X}}_t = D_t(\mathbf{Z}_t)$, 
`$\cdot$' denotes the element-wise product, and 
$\sigma$ is the sigmoid function that scales its input to $[0,1]$. We can similarly define $L_{roll}$ and $L_p$.

In each training epoch, we optimize \emph{both} the encoders and decoders by minimizing $L_{roll}$, $L_t$ and $L_p$ for the given training batch.
We refer to the way we train the model as using the ``temporal supervision,'' since to minimize the loss terms $L_{roll}$, $L_t$ and $L_p$, we have to make accurate prediction for each of the $T$ time frames.

When the adversarial training strategy is employed (i.e., those marked by dashed lines in Figure \ref{fig: model}a), we additionally consider the following two loss terms:
\begin{equation} 
L_t^n = - \textstyle{\sum} ~[\ln(1-\sigma(\widehat{\mathbf{X}}_t^n)) ] \,, 
\label{eq: BNC adv timbre}
\end{equation}
\begin{equation} 
L_p^n =  - \textstyle{\sum} ~[ \ln(1-\sigma(\widehat{\mathbf{X}}_p^n)) ] \,, 
\label{eq: BNC adv pitch}
\end{equation}
where $\widehat{\mathbf{X}}_t^n = D_t(\mathbf{Z}_p)$, $\widehat{\mathbf{X}}_p^n = D_p(\mathbf{Z}_t)$, meaning that we feed the `wrong' input (purposefully) to $D_t$ and $D_p$.

The equation must close to zero, which means when we use the wrong input, we expect $D_t$ and $D_p$ can output nothing (i.e., all zeros), since, e.g., $\mathbf{Z}_t$ is supposed not to contain any pitch-related information.

Please note that, in adversarial training, we use $L_t^n$ and $L_p^n$ to update the \emph{encoders only}. This is to preserve the function of the decoders in making accurate predictions.

\subsection{The UnetED Model}
The architecture of UnetED is depicted in Figure \ref{fig: model}b, which has a U-shape similar to \cite{ronneberger2015u}. In UnetED, we learn only one encoder $E_{cqt}$ to get a single latent representation $\mathbf{Z}_t$ of the input $\mathbf{X}_{cqt}$. We add skip connections between $E_{cqt}$ and $D_{roll}$, which enables lower-layer information of E (closer to the input) to be passed directly to the higher-layer of D (closer to the output), making it easier to train deeper models. The model learn $E_{cqt}$ and $D_{roll}$ by minimizing $L_{roll}$, the cross entropy between $D_{roll}(\mathbf{Z}_t)$ and the pianoroll $\mathbf{X}_{roll}$. Moreover, we promote timbre information in $\mathbf{Z}_t$ by refining $E_{cqt}$ and learning a classifier $D_t$ by minimizing $L_t$ (see Eq.~(\ref{eq: BCN})). When the adversarial training strategy is adopted, we design a GAN-like structure (i.e., unlike the case in DuoED) to dispel pitch information from timbre representation. That is, the model additionally learns a pitch classifier $D_p$ to minimize the loss $L_{p}$ between $D_p(\mathbf{Z}_t)$ and $X_p$. This loss function only updates $D_p$. Meanwhile, the encoder $E_{cqt}$ tries to fool $D_p$, where the ground truth matrices should be zero matrices. That is, the encoder should minimize $L_p^n$, and the loss function only affects the encoder.

The design of UnetED is based on the following  intuitions. First, since $\mathbf{Z}_t$ is supposed not to carry pitch information, the only way to obtain the pitch information needed to predict $\mathbf{X}_{roll}$ is from the skip connections. Second, it is fine to assume that the skip connections can pass over the pitch information, due to the nice one-to-one time-frequency correspondence between $\mathbf{X}_{cqt}$ and each frontal slice of $\mathbf{X}_{roll}$.\footnote{That is, both $X_{cqt}(i,j)$ and $X_{roll}(i,j,m)$ refer to the activity of the same musical note $i$ for the same time frame $j$.} Moreover, in $\mathbf{X}_{cqt}$ pitch only affects the lowest partial of a harmonic series created by a musical note, while timbre affects all the partials.
If we view pitch as the ``boundary'' outlining an object, and timbre as the ``texture'' of that object, detecting the pitches may be analogous to image segmentation, for which U-nets have been shown effective in solving \cite{ronneberger2015u}.

We note that the major difference between DuoED and UnetED is that there is a pitch code $\mathbf{Z}_p$ in DuoED. Although the pitch representation may benefit other tasks, for composition style transfer we care more about the timbre code $\mathbf{Z}_t$. 

\subsection{Composition Style Transfer}
\label{sec:method_transfer}
While both DuoED and UnetED are not trained to perform music rearrangement, they can be applied to music rearrangement due to the built-in pitch/timbre disentanglement. Specifically, for style transfer, we are given a source clip \texttt{A} and a target clip \texttt{B}, both audio files. We can realize style transfer by using \texttt{A} as the input to the encoder of DuoED or UnetED to get content information (i.e., through the pitch code  $\mathbf{Z}_p^{\texttt{(A)}}$ or the skip connections), but then combine it with the timbre code $\mathbf{Z}_t^{\texttt{(B)}}$ obtained from \texttt{B} to generate an original pianoroll $\mathbf{X}_{roll}^{\texttt{(A)}\rightarrow\texttt{(B)}}$, from which we can create synthesized audio. 

\subsection{Implementation Details}
\label{sec:implementation_details}
Since the input and output are both matrices, we use convolutional layers in all the encoders and decoders of DuoED and UnetED. To accommodate input of variable length, we adopt a fully-convolutional design~\cite{oquab15localization}, meaning that we do not use pooling layers at all. 
All the encoders in our models are composed of four residual blocks. Each block contains three convolution layers and two batch normalization layers.  The decoder $D_{roll}$ has the same structure, but use transpose convolution layers to do the upsampling. There are in total twelve layers in both encoder and pianoroll decoder $D_{roll}$. For the pitch and timbre decoder, we use three transpose convolution layers to reconstruct the pitch roll and instrument roll from latent representation. Moreover, we use leaky ReLU as the activation function for all layers but the last one, where we use the sigmoid function. Both DuoED and UnetED are trained using stochastic gradient descend with momentum 0.9. The initial learning rate is set to 0.005.

We use the newest released MuseScore dataset \cite{hung2018multitask} to train the proposed models. This dataset contains 344,166 paired MIDI and MP3 files. Most MP3 files were synthesized from the MIDIs with the MuseScore synthesizer by the uploaders.
Hence, the audio and MIDIs are already time-aligned. We further ensure temporal alignment by using the method proposed by \cite{raffel2016optimizing}. We then convert the time-aligned MIDIs to pianorolls with the \texttt{pypianoroll} package~\cite{pypianoroll}. 
The dataset contains music of different genres and 128 different instrument categories as defined by the MIDI spec.

\begin{table*}
\centering
\begin{tabular}{|r|l|c c c c c|c|} \hline
Method&training set& Piano &Guitar&Violin&Cello&Flute&Avg\\
 \hline\hline
 \cite{Hung2018FramelevelIR}& MuseScore training split &
0.690&0.660&0.697&0.774&0.860&	0.736\\
\cite{liu2018weakly} & YouTube-8M 
&0.766&0.780&0.787&	0.755&0.708&0.759\\
\cite{hung2018multitask} & MuseScore training split & 0.718&0.819&0.682&0.812&\textbf{0.961}&0.798\\
\cite{gururani18ismir}&M\&M training split &0.733&0.783&0.857&\textbf{0.860}&0.851&0.817\\\hline
DuoED updated&MuseScore (pre), M\&M training split&0.721&0.790&0.865&0.810& 0.912&0.815\\
UnetED updated&MuseScore (pre), M\&M training split&\textbf{0.781}&\textbf{0.835}&\textbf{0.885}&0.807& 0.832&\textbf{0.829}\\
\hline
\cite{hadad2018two} updated&MuseScore (pre), M\&M training split&0.745&0.807&0.816&0.769&0.883&0.804\\
\cite{liu2018multi} updated&MuseScore (pre), M\&M training split&0.808&0.844&0.789&0.766&0.710&0.793\\
\hline
\end{tabular}
\caption{Average AUC scores of per-second instrument activity detection on the test split of the `MedleyDB+Mixing Secret' (M\&M) dataset \protect\cite{Gururani2017MixingS,hung2018multitask}, for five instruments. 
For the latter four methods, we pre-train (`pre') the models on the MuseScore dataset \protect\cite{hung2018multitask} and then use the training split of M\&M for updating the associated timbre classifiers.}
\label{tab:inst_result_com}
\end{table*}

\begin{table}
\centering
\begin{tabular}{|r|l|c|}\hline
evaluated roll& model&Avg AUC\\
 \hline\hline
\multirow{4}{*}{Instrument roll}
&DuoED w/o Adv&0.731\\
&DuoED w Adv&\textbf{0.741}\\
&UnetED w/o Adv & 0.733\\
&UnetED w Adv &\textbf{0.754} \\
\hline
\multirow{4}{*}{Sum up from Pianoroll}
&DuoED w/o Adv&0.778\\
&DuoED w Adv&\textbf{0.781}\\
&UnetED w/o Adv & 0.778\\
&UnetED w Adv &\textbf{0.783} \\
\hline
\end{tabular}
\caption{The same evaluation task as that in Table \ref{tab:inst_result_com}. 
We compare the models trained on MuseScore only here, with (`w/') or without (`w/o') adversarial training (`Adv'). We evaluate both the instrument roll predicted by the timbre decoder $D_t$, and the instrument roll obtained by summing up the pianoroll predicted by $D_{roll}$.}
\label{tab:inst_result_abla}
\end{table}

\section{Experiment}
\label{sec:exp}
As automatic music rearrangement remains a new task, there are no standard metrics to evaluate our models.
We propose to firstly evaluate our models objectively with the surrogate task of \emph{instrument activity detection} (IAD) \cite{Gururani2017MixingS}---i.e., detecting the activity of different instruments for each short-time audio segment---to validate the effectiveness of the obtained instrument codes $\mathbf{Z}_t$. 
After that, we evaluate the performance of music rearrangement subjectively, by means of a user study.

\subsection{Evaluation on Instrument Activity Detection}
\label{sec:exp1}

We evaluate the performance for IAD using the `MedleyDB + Mixing Secret' (M\&M) dataset proposed by \cite{Gururani2017MixingS}.
The dataset contains real-world music recordings (i.e., not synthesized ones) of various genres. 
By evaluating our model on this dataset, we can compare the result with quite a few recent work on IAD. Following the setup of \cite{hung2018multitask}, we evaluate the result for per-second IAD in terms of the area under the curve (AUC). We compute AUC for each instrument and report the per-instrument AUC as well as the average AUC across the instruments.

IAD is essentially concerned with the prediction of the instrument roll shown in Figure \ref{fig: inst_roll}.  As our DuoED and UnetED models are pre-trained on a synthetic dataset MuseScore, we train additional instrument classifiers $D_t'$ with the pre-defined training split of the M\&M dataset (200 songs)  \cite{hung2018multitask}.  $D_t'$ takes as input the timbre code $\mathbf{Z}_t$ computed by the pre-trained models.
Following \cite{gururani18ismir}, we use a network of four convolution layers and two fully-connected layers for the instrument classifiers $D_t'$.\footnote{We do not use the M\&M training split to fine-tune the original instrument classifiers $D_t$ of our models but train a new one $D_t'$, so as to make sure that our instrument classifier has the same architecture as that used by  \cite{gururani18ismir}. 
In this way, the major difference between our models and theirs is the input feature (the timbre code $\mathbf{Z}_t$ for ours, and the log mel-spectrogram for theirs), despite of some minor differences in for example the employed filter sizes.}
We use the estimate of $D_t'$ as the predicted instrument roll.


Table \ref{tab:inst_result_com} shows the evaluation result on the pre-defined test split of the M\&M dataset (69 songs) \cite{hung2018multitask} of four state-of-the-art models (i.e., the first four rows) and our models (the middle two rows), considering only the five most popular instruments as \cite{hung2018multitask}.
Table \ref{tab:inst_result_com} shows that the proposed UnetED model can achieve better AUC scores in most instruments and achieve the highest average AUC 0.829 overall, while the performance of DuoED is on par with \cite{gururani18ismir}. We also conduct a paired t-test and found that the performance difference between Gururani’s method and the proposed UnetED method is statistically significant (p-value$<$0.05).
This result demonstrates the effectiveness of the learned disentangled timbre representation $\mathbf{Z}_t$ for IAD, compared to conventional representations such as the log mel-spectrogram used by \cite{gururani18ismir}.
Moreover, we note that these prior arts on IAD cannot work for composition style transfer, while our models can.

The last two rows of Table \ref{tab:inst_result_com} show the result of two existing disentanglement methods originally developed for images \cite{liu2018multi,hadad2018two}.\footnote{We adapt the two methods for music as follows. We employ the pianoroll as the target output for both. And, for \cite{liu2018multi}, we treat ‘S’ as timbre and ‘Z’ as pitch; for \cite{hadad2018two}, we replace ‘style’ with pitch and ‘class’ with timbre. } 
We use encoder/decoder structures similar to our models and the same strategy to  pre-train on MuseScore and then optimize the timbre classifier on the M\&M training split. We see that the proposes models still outperform these two models.
This can be expected as their models were not designed for music.

\begin{figure}
\centering
\includegraphics[width=\columnwidth]{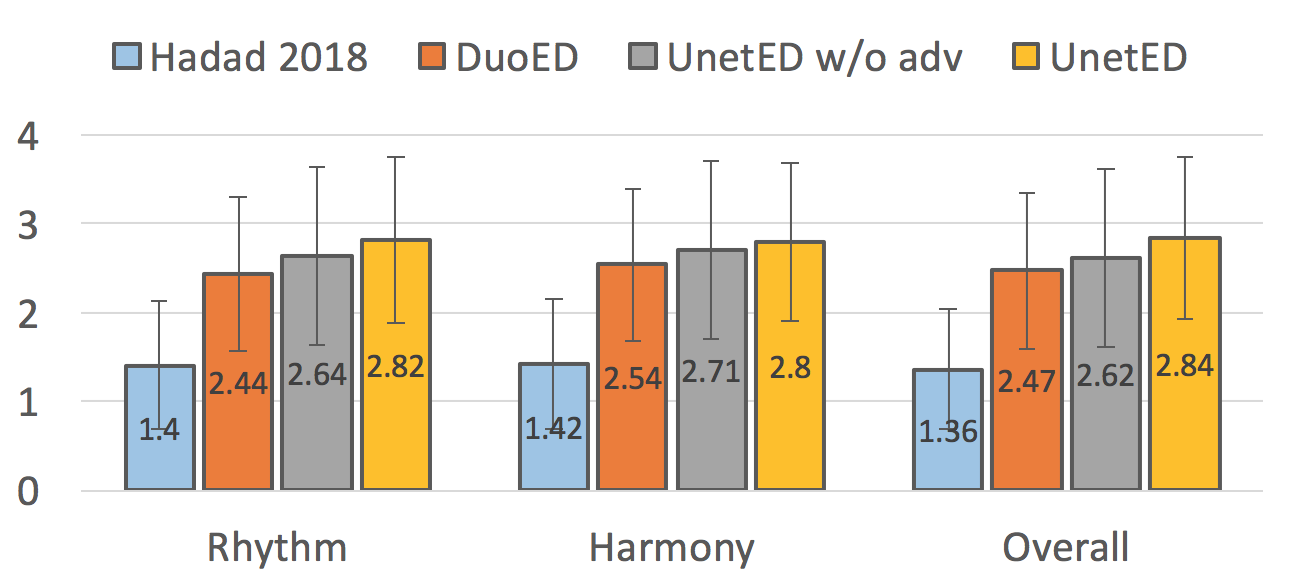}
\caption{Result of our user study on music rearrangement.}
\label{fig: userstudy}
\end{figure}

Table \ref{tab:inst_result_abla} reports an ablation study where we do not use the M\&M training split to learn $D_t'$ but use the original $D_t$ trained on MuseScore for IAD. In addition, we compare the case with or without adversarial training,
and the case where we get the instrument roll estimate from the output of $D_{roll}$ \cite{hung2018multitask}, which considers both pitch and timbre.
We see that adversarial training improves IAD, and that adding pitch information helps. From Tables \ref{tab:inst_result_com} and \ref{tab:inst_result_abla} we also see that it is helpful to update the instrument classifier by the training split of M\&M, possibly due to the acoustic difference between synthetic and real-world audio recordings.

\begin{figure*}[!htp]
\centering
\begin{subfigure}[c]{0.49\linewidth}
\includegraphics[width=\linewidth]{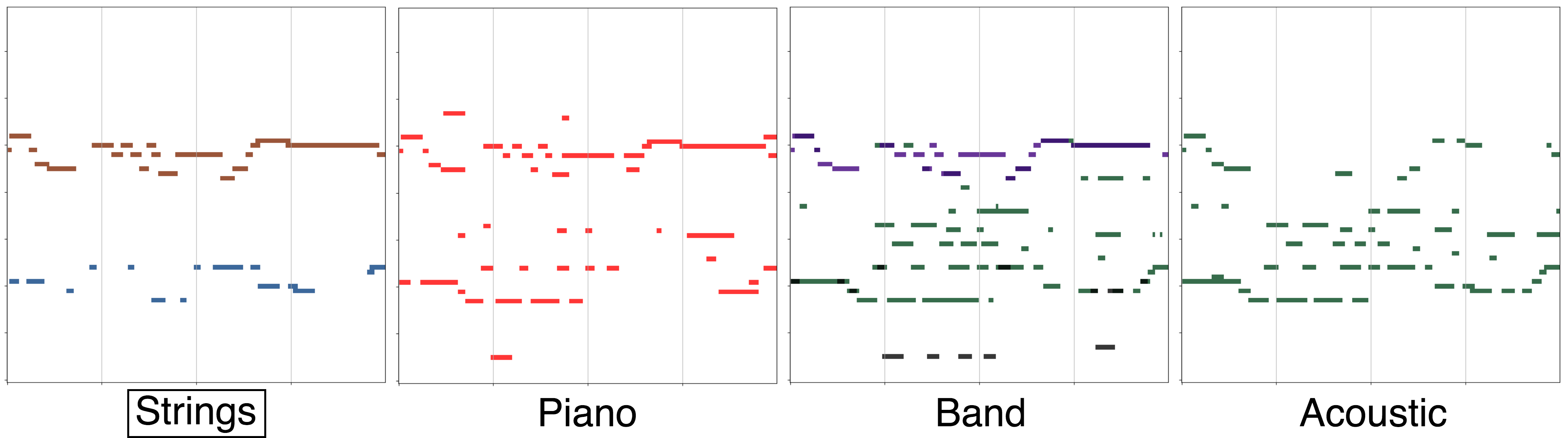}
\label{fig: string_result}
\end{subfigure}
\begin{subfigure}[c]{0.49\linewidth}
\includegraphics[width=\linewidth]{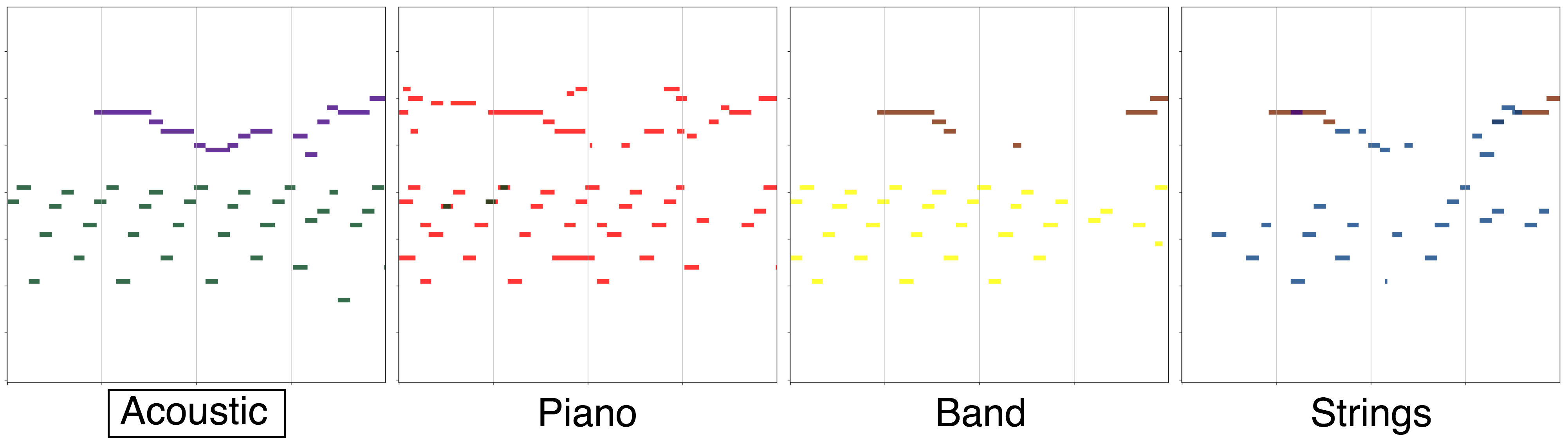}
\label{fig: acoustic_result}
\end{subfigure}
\begin{subfigure}[c]{0.49\linewidth}
\includegraphics[width=\linewidth]{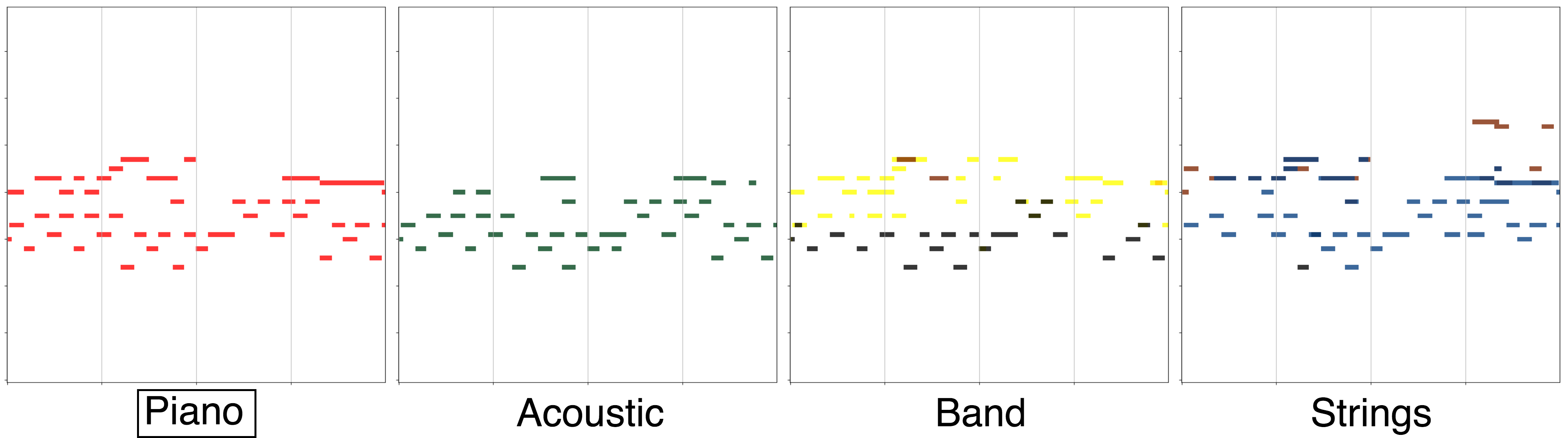}
\label{fig: piano_result}
\end{subfigure}
\begin{subfigure}[c]{0.49\linewidth}
\includegraphics[width=\linewidth]{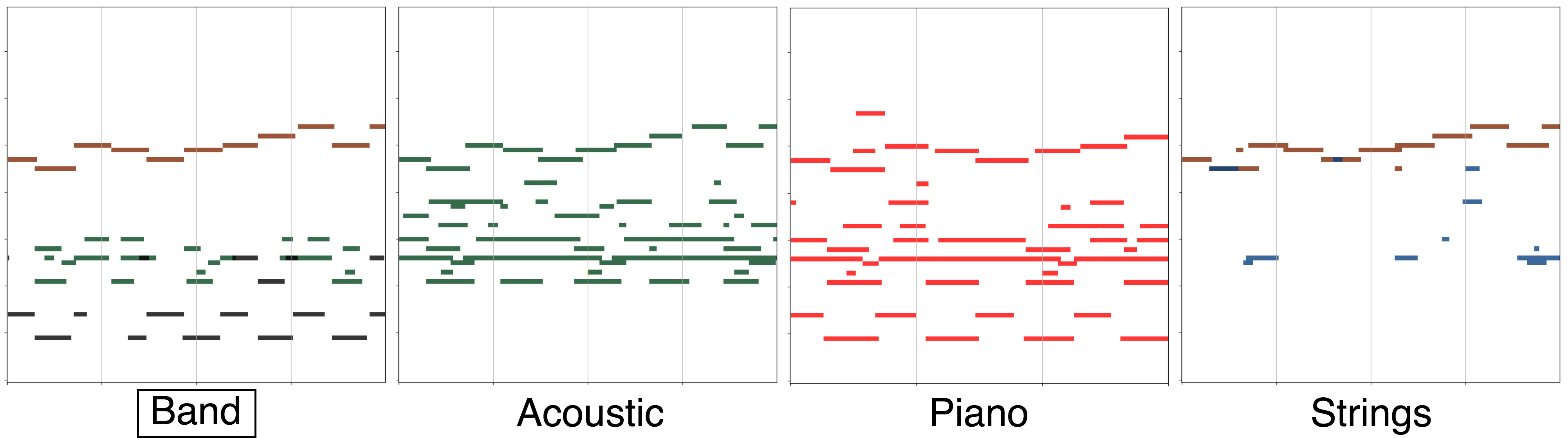}
\label{fig: band_result}
\end{subfigure}
\caption{Demonstration of music rearrangement by UnetED (best viewed in color), for four types of styles: strings, piano, acoustic, and band (see Section \ref{sec:exp2} for definitions). The source clips are those marked by bounding boxes, and the generated ones are those to the right of them.
[\underline{Purple}: flute, \underline{Red}: piano, \underline{Black}: bass, \underline{Green}: acoustic guitar, 
\underline{Yello}: electric guitar, \underline{Blue}: cello, \underline{Brown}: violin].} 
\label{fig: transfer_result}
\end{figure*}

\subsection{Evaluation on Music Rearrangement}
\label{sec:exp2}
To evaluate the result of music rearrangement, we choose four types of popular composition style in our experiment: \textbf{strings composition} (i.e., violin and cello), \textbf{piano}, \textbf{acoustic composition} (i.e., guitar, or guitar and melody), and \textbf{band composition} (i.e., electric guitar, bass, and melody); the melody can be played by any instrument.
We randomly choose four clips in each type from the MuseScore dataset, and then transfer them into the other three types following the approach described in Section \ref{sec:method_transfer}. 
We intend to compare the result of UnetDE, DuoDE, UnetDE w/o Adv, and  
the method proposed by \cite{hadad2018two}. We treat the last method as the baseline, as it is designed for image style transfer, not for music.

We invite human subjects to listen to and rate the rearrangement result. Each time, a subject is given the source clip, and the four rearranged results by different models, all aiming to convert the style of that clip to one of the other three. 
The subject rate the rearranged results in the following three dimensions in a four-point Likert scale:
\begin{itemize}
\renewcommand\labelitemi{--}
    \item whether the composition sounds \emph{rhythmic}; 
    \item whether the composition sounds \emph{harmonic};
    \item and, the \emph{overall quality} of the rearrangement.
\end{itemize}
The  scores are the higher the better. Since there is no ground truth of music style transfer, the rating task is by nature subjective. This process is repeated $4\times 3=12$ times until all the source-target pairs are evaluated. Each time, the ordering of the result of the four modes are random.

We distributed the online survey through emails and social media to solicit voluntary, non-paid participation.
40 subjects participated in the study, 18 of which have experience in composing music. 
Figure \ref{fig: userstudy} shows the average result across the clips and the subjects. 
It shows that, while DuoED and UnetED perform similarly for IAD, UnetED performs better in all the three metrics for music rearrangement.  
Moreover, with the adversarial training, the UnetED can generate more rhythmic and harmonic music than its ablated version. 
And, \cite{hadad2018two} does not work well at all.\footnote{The poor result of \cite{hadad2018two} can be expected, since the pre-trained instrument decoder can only guarantee $Z_t$ to contain timbre information, but cannot restrict $Z_t$ not to have any pitch information.
We observe that this method tend to lose much pitch information in the rearranged result and create a lot of empty tracks.
} Moreover, we found that there is no much difference between the ratings from people with music background and people without music background. The only difference is that people with music background averagely tends to give slightly lower ratings.

Figures \ref{fig: transfer_result} 
shows examples of the rearrangement result by UnetED. We can see that when rearranging the music to be played by band, UnetED finds the low-pitched notes for the bass to play and the melody notes for the flute or violin to play. This demonstrates that UnetED has the ability to choose suitable instruments playing certain notes combination. This result is also showed in arranging string instruments.
More demo results of UnetED can be found at \url{https://biboamy.github.io/disentangle_demo/.} 


\section{Conclusion}
\label{conclusion}
In this paper, we have presented two models for learning disentangled timbre representations. This is done by using the instrument and pitch labels from MIDI files. Adversarial training is also employed to disentangle the timbre-related information from the pitch-related one. Experiment show that the learned timbre representations lead to state-of-the-art accuracy in instrument activity detection.
Furthermore, by modifying the timbre representations, we can generate new score rearrangement from audio input. 

In the future, we plan to extend the instrument categories and also include the singing voice to cope with more music genres. We want to test other combinations of instruments to evaluate the performance of our models. 
We also want to further improve our models by re-synthesizing the MP3 files of the MuseScore data with more realistic sound fonts. 

\bibliographystyle{ijcai19}
\bibliography{ijcai19}
\end{document}